\documentclass[traditabstract]{aa} 
\usepackage{graphicx}
\usepackage{txfonts}
\begin{document}
   \title{An equatorial ultra iron-poor star identified in BOSS}
   \author{C. Allende Prieto
          \inst{1} \inst{2}
          \and
          E. Fern\'andez-Alvar\inst{1} \inst{2}
      \and 
	  D. S. Aguado\inst{1} \inst{2}
      \and
          J. I. Gonz\'alez Hern\'andez\inst{1} \inst{2}	  
      \and
          R. Rebolo\inst{1} \inst{2} \inst{3}
      \and
          Y. S. Lee\inst{4}
      \and
          T. C. Beers\inst{5}
      \and
         C. M. Rockosi\inst{6}
      \and 
         J. Ge\inst{7}
          }

   \institute{Instituto de Astrof\'{\i}sica de Canarias,
              V\'{\i}a L\'actea, 38205 La Laguna, Tenerife, Spain\\              
         \and
             Universidad de La Laguna, Departamento de Astrof\'{\i}sica, 
             38206 La Laguna, Tenerife, Spain \\      
         \and
               Consejo Superior de Investigaciones Cient\'{\i}ficas, Spain      \\
         \and
               Department of Astronomy and Space Science, Chungnam National University, Daejeon 305-764, Republic of Korea \\
         \and
       Dept. of Physics and JINA Center for the Evolution of the Elements, University of Notre Dame, Notre Dame, IN 46556, USA \\
         \and
             UCO/Lick Observatory, University of California, Santa Cruz, CA 95064, USA \\
         \and
 		Department of Astronomy, University of Florida, Bryant Space Science Center, Gainesville, FL 32611-2055, USA \\
%
             }

   \date{Received February 2015; accepted xxxx}

 

\abstract{
We report the discovery of SDSS J$131326.89-001941.4$, an ultra iron-poor red
giant star
([Fe/H]$\simeq -4.3$) with a very large carbon abundance ([C/Fe]$\simeq +2.5$).
This object is the fifth star in this rare class, and the combination 
of a fairly low effective temperature ($T_{\rm eff} \simeq 5300$ K), which  
enhances line absorption,   with its brightness ($g=16.9$), makes it possible to 
measure the abundances of calcium, carbon and iron using a low-resolution 
spectrum from the Sloan Digital Sky Survey. We examine the carbon and
iron abundance ratios in this star and other similar objects in the
light of predicted yields from metal-free massive stars, and conclude that they
are consistent. By way of comparison,  stars with similarly low iron 
abundances but lower carbon-to-iron ratios deviate from the theoretical
predictions.  
}


   \maketitle
%

\section{Introduction}

Only hydrogen, helium and traces of lithium were produced in the primordial nucleosynthesis,
shortly after the Big Bang. The first generation of stars to form must have had 
the same primordial mixture of nuclei. These Population III stars 
started the reionization of the 
universe, and created the first nuclei heavier than lithium (Truran \& Cameron 
1971, Silk 1997). 

At zero metallicity, gas cooling is difficult, and the formation
of massive stars, with $M>>1 M_{\odot}$, 
is favored (Bromm, Coppi \& Larson 1999; Abel \& Norman 2002). 
Once the metallicity of the interstellar medium reaches a level of
about $10^{-4}$ times solar, cooling by fine structure lines of atomic
oxygen and ionized carbon 
makes it much easier to form low-mass stars (Bromm \& Loeb 2003).
Almost all of the stars with the lowest iron abundances known ([Fe/H]$<-4.5$) 
exhibit extreme carbon
enhancements, which explains their formation despite their low mass 
(Christlieb et al. 2002; Frebel et al. 2005; Norris et al. 2007; Keller et al. 2014).
However, the star recently identified by Caffau et al. (2011, 2012) with an
iron abundance [Fe/H]$\simeq -4.9$, exhibits a much lower carbon abundance,
suggesting that earlier predictions of the initial mass function for zero-metallicity
stars need to be seriously revised 
(Scheneider et al. 2012; Clark et al. 2013; 
Greif et al. 2013; Ji et al. 2014). 

The chemical patterns observed in the atmospheres of ultra-low metallicity stars 
not only illuminate the investigation of star formation at low metallicity,
but, since such stars formed out of gas that may have been polluted by one
or very few supernovae (Audouze \& Silk 1995), 
they can constrain the nucleosynthetic yields of the supernovae
that ended the lives of the first stars. As of this writing, decades of search
have only identified about a dozen stars with [Fe/H]$<-4$ (see Section \ref{conclusions}). 
In this paper,
we report on the discovery of an additional specimen in this category, 
currently one of the tenth most iron-poor stars known.

\section{Observations}
\label{observationat}

We identified SDSS J$131326.89-001941.4$ (hereafter J1313$-0019$) in observations 
of the Baryonic Oscillations Spectroscopic Survey
(BOSS; Eisenstein et al. 2011; Dawson et al. 2013) 
obtained in March 2014 with the Sloan 2.5m telescope (Gunn et al. 2006) and
its optical spectrographs (Smee et al. 2013), 
and released in DR12, the final data release
of the Sloan Digital Sky Survey III (SDSS; Alam et al. 2015). The BOSS 
spectrum of this star, bright for SDSS standards 
($u=17.92, g=16.87, r=16.41, i=16.18, z=16.10$),
was obtained
in the next-to-last plate of the survey. This plate (7456) and a few others,
were targeted with the purpose of cross-calibrating BOSS stellar
observations with those from the Sloan Extension for Galactic
Understanding and Exploration (SEGUE; Yanny et al. 2009), obtained with the previous
version of the SDSS spectrographs, and the Gaia-ESO survey (Gilmore et al. 2012).  

The star had been observed previously as 
part of SEGUE (plug-plate 2901), but that spectrum was of a lower quality than
the BOSS spectrum, with a median signal-to-noise ratio per 
pixel\footnote{A pixel in the extracted and resampled SDSS spectra    
spans 69 km s$^{-1}$ or 1 \AA\ at 4500 \AA.} between 5000 and 8000 \AA\ of about 40. 
The BOSS spectrum has an average signal-to-noise
ratio per pixel of about 60, and enables a fairly detailed analysis.

J1313$-0019$ was targeted by SEGUE as a metal-poor star candidate. In addition
to having a lower signal-to-noise ratio, the SEGUE spectrum shows some systematic
residuals that made it difficult to recognize the target as an ultra metal-poor
star.

\section{Analysis}
\label{analysis}

We analyzed the BOSS spectrum of J1313$-0019$ with the software and models described
by Allende Prieto et al. (2014). Briefly, the entire spectrum between 
385-919 nm is matched with model spectra based on 
classical model atmospheres. We made use of the {\tt FERRE}\footnote{{\tt FERRE}
is available from http://hebe.as.utexas.edu/ferre.} code (Allende Prieto
et al. 2006), which searches for the
model parameters that best match the observations by interpolating in a 
pre-computed grid of synthetic spectra. 
The analysis, which assumed a micro-turbulence velocity of 2 km s$^{-1}$,
 indicated $T_{\rm eff} = 5251 \pm 50$ K , $\log g = 1.1 \pm 0.2$, and 
[Fe/H]$=-4.5 \pm 0.1$, where the error bars combine in quadrature the 
internal uncertainties of the code with those inferred from the comparison with
the SEGUE Stellar Parameter Pipeline (SSPP; Lee et al. 2007 and later updates) 
made by Allende Prieto et al. (2014).
From the analysis of the globular cluster M13 ([Fe/H]$\simeq -1.7$), 
Allende Prieto et al. detected
significant systematic offsets in their inferred parameters for 
metal-poor low-gravity ($\log g < 3$) stars.
Taking those offsets into account, the preferred parameters for J1313$-0019$ would be
$T_{\rm eff} = 5320 \pm 150$ K , $\log g = 2 \pm 0.5$, and   
[Fe/H]$=-4.2 \pm 0.2$. This analysis adopted a heliocentric radial velocity of 
$242 \pm 4$ km s$^{-1}$, derived by the SDSS {\it spec1d} pipeline (Bolton et al. 2012).  

The SEGUE spectrum of J1313$-0019$,
although of a significantly worse quality than the BOSS spectrum and
afflicted by some obvious artifacts, 
suggests $T_{\rm eff} = 5505 \pm 60$ K , $\log g = 3.9 \pm 0.3$, and   
[Fe/H]$=-4.2 \pm 0.1$. This spectrum was also analyzed by the SSPP  
which arrived at the following parameters: 
 $T_{\rm eff} = 5674 \pm 122$ K , $\log g = 2.41 \pm 0.05$, and   
[Fe/H]$=-4.25 \pm 0.08$\footnote{We adopt the values released in the ninth data release
of the SDSS (DR9; Ahn et al. 2012).}. From this spectrum, the {\it spec1d} pipeline measured a 
heliocentric radial
velocity of $268 \pm 6$ km s$^{-1}$, and provided a second value of $269 \pm 4$ km s$^{-1}$
from the fitting
of the data with smoothed templates from the Elodie library in the 4100--6800 \AA\ 
range. This velocity is different from the one measured from the BOSS spectrum 
by 26 km s$^{-1}$ or about 6 $\sigma$, suggesting the star may be part of a binary 
system, but more data are necessary to confirm this hypothesis.

\begin{figure*}
   \centering
   \includegraphics[width=13cm,angle=0]{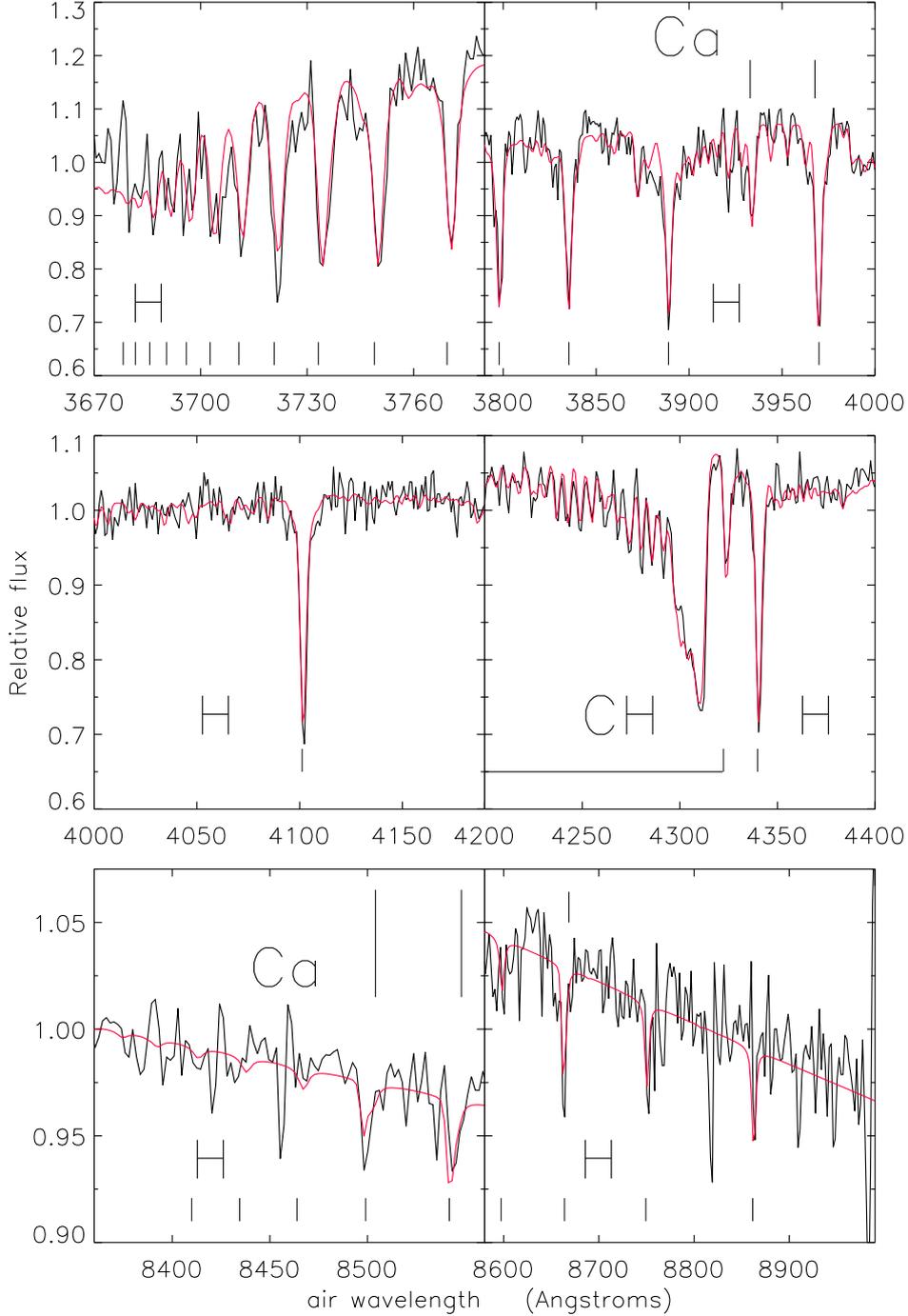}
      \caption{BOSS spectrum of J1313$-0019$ (black) and best-fitting model (red).
The elements or molecules associated with the most prominent lines are 
identified.
The continuum correction
has been carried out following the procedure described by Allende Prieto
et al. (2014), namely dividing the spectral range in 20 segments and
dividing the flux in each by its mean value. This algorithm is very
robust to changes in the signal-to-noise ratio, and our own simulations
show that has excellent performance.}
         \label{f1}
   \end{figure*}

The extinction estimate in the Schlegel et al. maps in the direction
of J1313$-0019$ is E(B-V)=0.028 mag. Correcting the observed $(g-r)$ color
of the star, adopting $A_g=1.19 A_V$ and $Ar=0.87 A_V$ (Girardi et al. 2004) and
the average Galactic relationship $A_V=3.1 E(B-V)$, we find $(g-r)_0=0.49$. 
Using the polynomial relationship proposed by Pinsonneault et al. (2012), we arrive
at $T_{\rm eff}=5576 $ K. On the other hand,
the SSPP reports an IRFM-based temperature of $5669 \pm 58$ K. Both are significantly
warmer than the temperatures we derive spectroscopically from the BOSS spectrum, 
but closer to those from the lower quality data from SEGUE. Nevertheless, we have 
confidence in the effective temperature inferred from the BOSS spectrum,
about 5400 K, since it reproduces both the local continuum slope and
the strength of the H lines in the spectrum 
 (see Fig. \ref{f1}).

The spectrum of J1313$-0019$ shows strong CH absorption, indicating that the star
is rich in carbon. The most isolated strong Fe I lines in the spectrum are in the blends 
around 3820 and 3860 \AA,
plus several fairly strong transitions between 3600 and 3608 \AA, but the
signal-to-noise ratio of the BOSS spectrum does not allow an unambiguous
detection of individual lines or groups of overlapping iron lines. 
The overall effect of Fe I absorption between 3600 and 3900 \AA, however,
is fairly significant, with dozens of features that reduce the stellar flux
by a few percent. These, as a whole, can be studied with  a statistical approach. 
We attempted to simultaneously recover 
the abundances of carbon, the $\alpha$-elements
(O, Mg, Si, S, Ca and Ti), and iron (and the rest of the metals), 
assuming again a micro-turbulence velocity of 2 km s$^{-1}$, and using 
(as in previous cases) $\chi^2$ as a metric. 
In this  analysis, which used the full BOSS spectral range (360-1000 nm)
 we arrived at
$T_{\rm eff} = 5378 \pm 50$ K , $\log g = 3.0 \pm 0.2$, [C/Fe]$=+2.5 \pm 0.1$,
[$\alpha$/Fe]$=+0.2 \pm 0.1$ and  
[Fe/H]$=-4.3 \pm 0.1$.
We also processed the BOSS spectrum of J1313$-0019$ through a customized version 
of the SSPP in which algorithms that used photometric information were
not used, arriving at $T_{\rm eff}= 5502 \pm 62$ K, $\log g= 2.69 \pm 0.41$, 
[Fe/H]$=-4.37 \pm 0.15$ and [C/Fe]$=2.68 \pm 0.10$.

\begin{figure}
   \centering
   \includegraphics[width=9cm,angle=0]{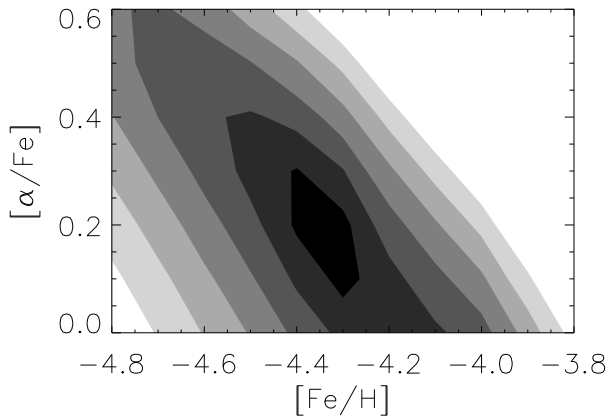}
      \caption{Contours of the $\chi^2$ statistic in the [Fe/H]-[$\alpha$/Fe] plane, corresponding to 
	1-2-3...etc $\sigma$ uncertainties in these parameters. }
         \label{f2}
   \end{figure}

Fig. \ref{f1} illustrates how well the best-fit model matches the observed BOSS spectrum
over several relevant spectral windows, including the 
wavelengths of the Ca II resonance lines, the G band (CH), 
and the Ca II infrared triplet. 
The H Balmer lines are the most prominent transitions 
in the top panels, and at 4102 (H$\delta$) and 4341 \AA\ (H$\gamma$) in the middle panels.
Molecular CH absorption is responsible for most of the remaining features apparent 
between 3850 and 4400 \AA, with the notable exception of the Ca II K line
at 3934 \AA\ (the Ca II H line overlaps with H$\epsilon$ at 3969 \AA).
The strong transitions in the bottom panels correspond to the
Ca II triplet (8498, 8542 and 8662 \AA) and (weaker) H Paschen lines at 8408, 8432, 
8461, 8497, 8540, 8593, 8659, 8744, and 8857  \AA.
Fig. \ref{f2} shows the chi-square contours 
in the [$\alpha$/Fe]-[Fe/H] plane in the
vicinity of the minimum. The contours correspond to values of the minimum plus
1, 4, 9, etc. corresponding to the 1-2-3-etc. $\sigma$ uncertainties if
we are dealing with a Gaussian error distribution (68\%, 95\%, 99.7\%, etc).
The reduced $\chi^2$ at the minimum is 1.1. 

We measured the equivalent width of the Ca II K line to be 0.64 \AA, and using
the July 2014 version of MOOG (Sneden 1971) and an appropriate ATLAS9 model atmosphere 
computed as in M\'esz\'aros et al. (2012), we found that such line strength 
correspons to [Ca/H]$=-4.0$. We note in passing that an interstellar 
component is apparent in the Ca II resonance lines, blue-shifted by about 
350 km s$^{-1}$, and which is apparent in the observation and not
the stellar model shown in the top right-hand panel of Fig. \ref{f1}.
The infrared Ca II lines are blended
with  Paschen lines, thus an equivalent-width analysis to 
estimate the Ca abundance would 
require modelling the contribution of the hydrogen lines 
(see Fig. \ref{f1}).
By adopting the atmospheric
parameters obtained from the global fit and minimizing $\chi^2$ over
the windows that are most sensitive to iron lines, as described in Fern\'andez-Alvar
et al. (2015), we arrive at [Fe/H]$=-4.4 \pm 0.3$. All of these checks
give results that are consistent with our previous analyses.

We do not directly detect lines of barium or magnesium; only 
upper limits can be given. 
The $\alpha$-element abundance is constrained mainly by 
the Ca II resonance lines at 3950 \AA\  and the Ca II infrared triplet at
8450--8600 \AA. 
Examining the spectrum in the vicinity of the Ba II resonance line at 4554 \AA\ 
we estimate [Ba/Fe]$<+1.5$,  and from an analysis of the region around the 
Mg I$b$ lines we estimate [Mg/Fe]$<+0.5$.

\section{Discussion and conclusions}
\label{conclusions}

J1313$-0019$ is a new example of the class of stars with extremely low 
iron abundances
and very high C/Fe ratios. This class includes about half of the stars
with [Fe/H]$<-4$. Previous high-resolution spectroscopic studies of these 
stars have shown that they are associated with the CEMP-no class 
(carbon-enhanced metal-poor stars without over-abundances of neutron-capture elements; 
see, e.g., Beers \& Christlieb 2005; Hansen et al. 2014). 
The relatively low effective temperature of this star ($T_{\rm eff}\sim 5400$ K)
and the high signal-to-noise of his BOSS spectrum, allows us to
determine the abundances of carbon, calcium and iron, despite the 
fairly low spectral resolution ($\lambda/\delta\lambda \sim 2000$). 
From our estimated calcium abundance [Ca/H]$\simeq -4.0$, and 
iron abundance [Fe/H]$\simeq -4.3$, J1313$-0019$ is among the ten 
most iron-poor stars known.

  \begin{figure}
   \centering
   \includegraphics[width=9cm,angle=0]{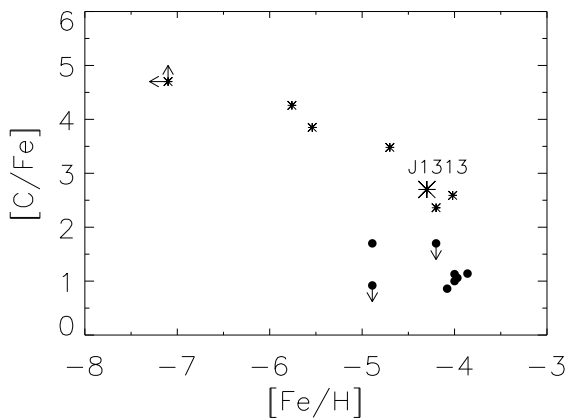}
      \caption{Carbon-to-iron ratio vs. iron abundance in stars with [Fe/H]$<-4$.  
Asterisks are used for those with [C/Fe]$>+2$, and filled circles for those
with [C/Fe]$<+2$. A larger symbol identifies the newly identified star J1313$-0019$.}
         \label{f3}
   \end{figure}

Fig. \ref{f3} shows the C/Fe ratio for the stars with the lowest iron abundances
known. In the case of SMSS J$031300.36-670839.3$ (Keller et al. 2014), only an upper
limit to the iron abundance is available ([Fe/H]$<-7.1$). The sample
of metal-poor stars known dramatically increases at [Fe/H]$>-4$. 
Chemical evolution models predict that the transition between a regime
dominated by stochastic star-formation events, where the gas is
polluted by a single or a few supernovae, and a well-mixed interestellar
gas where the  instantaneous mixing approximation  works, takes
place at about [Fe/H]$\sim -4$ (Audouze \& Silk 1995; Oey 2003; Karlsson 2005).

  \begin{figure*}
   \centering
   \includegraphics[width=13cm,angle=0]{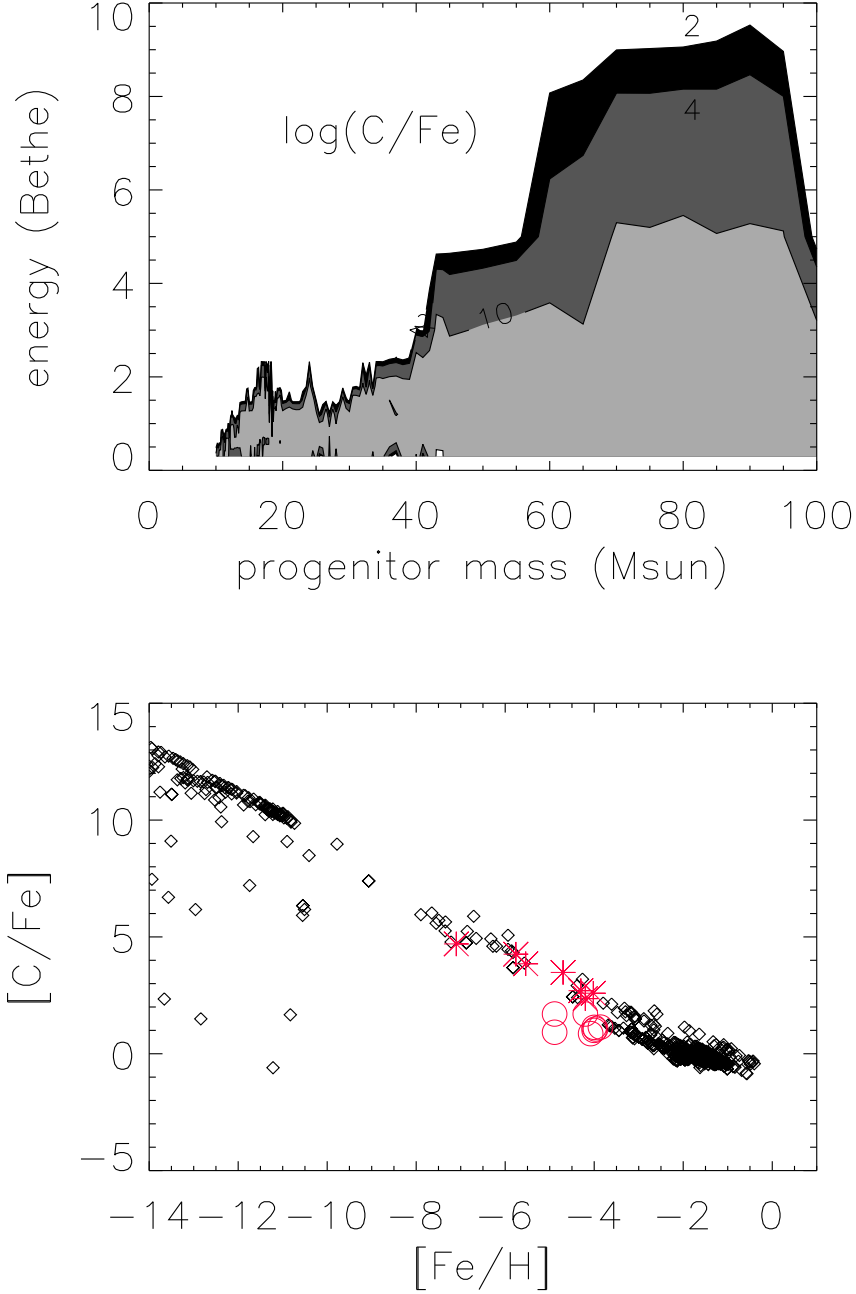}
      \caption{Top panel: Contour diagram showing the 
carbon-to-iron mass ratio, as a function of progenitor mass
and explosion energy, predicted for the supernovae yields of
zero metallicity stars by Heger \& Woosley (2010). Contours are shown for 
$\log$(C/Fe)$=2,4,10$ and 20.
Bottom panel: The carbon-to-iron ratio, as a function of iron production, predicted for
the supernovae models (rhombi; with iron multiplied by an arbitrary factor), and
abundances found in the most iron-poor stars known (the upper limits for
iron in SMSSJ$031300.36-670839.3$ have been ignored for simplicity).
The asterisks correspond to stars that have [C/Fe]$>+2$, and the open circles
show those with lower carbon-to-iron ratios.}
         \label{f4}
   \end{figure*}

 We focus our attention on the
13 stars known with [Fe/H]$<-4$. The parameters for these stars 
are drawn from those derived or compiled by  
Caffau et al. (2012), Norris et al. (2013), Keller et al. (2014), and Hansen et al. (2014).
About half of the sample, including
J1313$-0019$, have 
a very large carbon enrichment [C/Fe]$>+2$, while the others
exhibit a lower C/Fe ratio, but still supersolar.  The most carbon-rich
stars show that the C/Fe ratio
is anticorrelated with the iron abundance. This implies that, as
suggested by Spite et al. (2013), the
carbon abundance in these stars is [C/H]$\sim -1.5$, irrespective
of [Fe/H].
This carbon abundance is well above the threshold for 
line cooling that facilitates the gas cloud fragmentation required
to form low-mass stars (Frebel et al. 2009).

The source of carbon for ultra metal-poor stars 
is hydrostatic helium burning in high-mass zero-metal (population III)
stars that end their lives as supernovae. The amount of carbon
released in each event is a fairly sensitive function of the 
mixing and fallback of material in the supernovae, which is difficult to model
in detail.   
Keller et al. have recently argued that the abundance pattern 
in SMSS J$031300.36-670839.3$
could be created by a single supernova with a mass of the
progenitor metal-free star in the vicinity of 60 M$_{\odot}$.
We have explored yields from models by Heger and Woosley (2010), and
as shown in the upper panel of Fig. \ref{f4}, there are multiple
combinations of the progenitor mass and the explosion energy
that lead to C/Fe mass ratios in the range found in the carbon-rich
ultra iron-poor stars discussed in this paper. These models
 consider non-rotating stars with masses between 10 and 100 M$_{\odot}$,
and follow their evolution, and final supernova explosions. We note
that there are several other scenarios for the production of the
elemental abundances found in carbon-rich ultra metal-poor stars (see, 
e.g. Maeder et al. ), but
we examine these in detail since they have already been used in previous
studies of similar stars, and they are provided in a format that can be
readily used for comparison with observations. Metal-free
stars with masses between 140-260 M$_{\odot}$ end their lives as
pair-instability supernovae, but calculations by Heber \& Woosley (2002) show
that their carbon-to-iron ratios are close to solar, with the exception 
of the stars in the lowest end of that range, whose yields can reach up to a
mass fraction C/Fe$\sim 10^{10}$.

The lower panel of Fig. \ref{f4} shows the log of the ratio of the carbon and
iron yields, against the log of the iron yield in solar masses, again
following calculations by Heger \& Woosely (2010). 
The ejecta from the supernovae will be diluted in the interstellar
gas, but the carbon-to-iron ratio should in principle be preserved.
After changing from mass fraction to particle fraction, we  
have subtracted 0.94, the log of the solar ratio of carbon
and iron abundances, to the supernova yields before placing them
in the figure ([C/Fe] = log (C/Fe) + log (56/12) - 0.94).
The iron-to-hydrogen ratio in the interstellar
medium (and in the atmospheres of these primitive stars) can be
arbitrarily shifted from the supernova yields 
in solar masses to make the horizontal axis in the figure, 
and we have shifted the log of the iron yield in solar masses
by $-0.7$ dex in order to match the observations.

There is a fairly tight relationship between the carbon-to-iron
ratio and the iron yields for these supernovae. For some models 
the carbon-to-iron ratios are much larger, up to [C/Fe] of about 20 
and out of the range shown
in the figure, but few models predict
lower ratios, especially for high values of the iron yields. 
All of the stars with [Fe/H]$<-4$ and [C/Fe]$>+2$ (asterisks in Fig. \ref{f4}), 
including J1313$-0019$, match the trend predicted by theory, 
suggesting that those stars could be formed out of gas
polluted by supernovae resulting from metal-free stars in the range 10-100 M$_{\odot}$.
This correlation is not observed for the ultra metal-poor stars that
show [C/Fe]$<+2$, indicating that their chemical composition is associated
with a different type of source, that the models considered do not span
a sufficiently large range in explosion energy,  or a systematic effect exists due 
to rotation (missing in the models).
Interestingly, neither the iron yields, nor the carbon-to-iron ratios of the
models 
correlate with the  mass of the progenitor. There also does not exist 
a correlation of the abundances with the energy of the explosion. 

Our result of supernovae from Population III stars matching the compositions
of ultra iron-poor carbon-rich stars is in line with previous investigations 
that found that models can reproduce in a fairly good detail the chemical
patterns measured in such stars (e.g., Joggerst et al. 2010; 
Keller et al. 2014). One of our most surprising findings is that 
we are able to measure several elemental abundances for ultra-low metallicity 
stars from low-resolution spectroscopy. 
The literature is polarized between studies at high spectral resolution
and high signal-to-noise ratios, and those at a modest resolution
with limited signal-to-noise ratios. However, 
high signal-to-noise observations
at limited resolution seem to provide a promising avenue 
for making significant advances in this field.
The Sloan Digital Sky Survey continues to obtain observations 
of stars at high Galactic lattitude, and to identify chemically
primitive stars. Building a larger sample of such rare objects will
be critical for understanding the formation and early chemical evolution of
galaxies, and in particular the Milky Way.

\begin{acknowledgements}

T.C.B. acknowledges partial support for this work from grants PHY 08-22648: Physics
Frontier Center/Joint Institute for Nuclear Astrophysics (JINA), and PHY 14-30152: 
Physics Frontier Center/JINA Center for the Evolution of the Elements (JINA-CEE), 
awarded by the US National Science Foundation.

Funding for SDSS-III has been provided by the Alfred P. Sloan Foundation, 
the Participating Institutions, the National Science Foundation, and the U.S. 
Department of Energy Office of Science. The SDSS-III web site is http://www.sdss3.org/.

SDSS-III is managed by the Astrophysical Research Consortium for the Participating 
Institutions of the SDSS-III Collaboration including the University of Arizona, 
the Brazilian Participation Group, Brookhaven National Laboratory, 
Carnegie Mellon University, University of Florida, the French Participation Group, 
the German Participation Group, Harvard University, 
the Instituto de Astrofisica de Canarias, the Michigan State/Notre Dame/JINA 
Participation Group, Johns Hopkins University, Lawrence Berkeley National Laboratory, 
Max Planck Institute for Astrophysics, Max Planck Institute for Extraterrestrial Physics, 
New Mexico State University, New York University, Ohio State University, 
Pennsylvania State University, University of Portsmouth, Princeton University, 
the Spanish Participation Group, University of Tokyo, University of Utah, 
Vanderbilt University, University of Virginia, University of Washington, 
and Yale University.

\end{acknowledgements}

\end{document}